\documentclass[preprint,12pt,authoryear]{elsarticle}

\usepackage{epsfig}

\usepackage[authoryear]{natbib}

\usepackage{amssymb}

\usepackage{graphicx}

\journal{Planetary and Space Science}

\begin{document}

\begin{frontmatter}



\title{Near-Earth Asteroids of Cometary Origin Associated with the Virginid Complex}






\author[address]{G.I. Kokhirova\corref{mycorrespondingauthor}}
\cortext[mycorrespondingauthor]{Corresponding author}
\ead{Kokhirova2004@mail.ru}
\author[address]{ A.I. Zhonmuhammadi }
\author[address]{U.H. Khamroev}
\author[mymainaddress] { T.J. Jopek }
\address[address]{Institute of Astrophysics of the National Academy of Sciences of Tajikistan, Ayni 299/5, Dushanbe, 734063, Tajikistan}
\address[mymainaddress]{Astronomical Observatory Institute, Faculty of Physics, A. M. University, Poznan, Poland }
\begin{abstract}
The Virginid meteoroid streams produce a series of meteor showers active annually during February-May. A certain parent comet is not found but a related association of some showers with near-Earth asteroids was previously established and a cometary origin of these asteroids was suggested. We performed a new search for NEAs belonging to the Virginid asteroid-meteoroid complex. On the base of calculation of orbital evolution of a sample of NEAs and determination of theoretical features of related showers a search for observable active showers close to theoretically predicted ones was carried out. As a result, the predicted showers of 29 NEAs were identified with the showers of the Virginid complex. Revealed association points to a cometary nature of NEAs that are moving within the stream and may be considered as extinct fragments of a larger comet-progenitor of the Virginid asteroid-meteoroid complex.
\end{abstract}


\end{frontmatter}
\section{Introduction}
Besides of major planets a lot of smaller bodies are moving in the Solar System. This population consists of comets, asteroids, and meteoroids. Comets and asteroids are the parents of meteoroids. A huge number of meteoroids generated by a one parent body form a meteoroid stream. The stream's meteoroids are moving in the interplanetary space by similar orbits close to the orbit of the parent. It is accepted that the activity of comets or their destruction are responsible for a formation of meteoroid streams. Forming of the stable long-lived meteoroid stream can be provided by a periodic normal gas- and dust producing activity of comet which is observing during passage of the perihelion \citep{whi50,whi51,bre54}. Moreover, a stream might be generated as a result of the catastrophic decay of comet due to an impact or other processes \citep{gro07,wes20}.  However, a break-up of a comet or an asteroid cannot lead to a formation of meteoroid stream, at least, the long-lived stream, since in this case the ejection of dust and debris will be once that is not sufficient for a formation of a stable stream.   

 A concept of the formation of meteoroid streams as a result of the cometary activity, as well as the circumstances of their evolution and structure are described in a lot of papers \citep[see e.g.] []{whi50,whi51,hug86,bab92,wil93,bab08a,bab15a}. When the Earth crosses the meteoroid stream orbits, the meteor showers are occurring.  As was shown by \citet{bab92} one stream might produce from four till eight meteor showers observable on the Earth. The quadruple intersections are the most spread event for meteoroid streams. For instance, the Taurid meteoroid stream consists of four meteor showers. At the pre-perihelion crossing with the Earth's orbit they are the Northern and Southern Taurids (17/NTA and 2/STA according to the official nomenclature given in the IAU MDC, see \citet{jop11,jop14,jop17} and after the naming rules have been modified in \citet{jop23}), which are observable on the Earth annually in September-November, as well as the Daytime $beta$-Taurids (173/BTA) and Daytime $zeta$-Perseids (172/ZPE) observable in June-July which are occurring at the post-perihelion crossing. The well-known parent body of the Taurid stream is comet 2P/Encke; however, it turned out that else more than 40 near-Earth asteroids (NEAs) relate to this family \citep{ash93a,ash93b,bab01,bab08a,por04,por06,mad13,rud12a,rud12b,kok22}.  The related association of these objects was found and the family was called as the Taurid asteroid-meteoroid complex. It was defined that asteroids of the Taurid complex very likely are extinct cometary nuclei or dead fragments of the larger comet-progenitor \citep[see e.g.][]{ash93a,por06,bab08a}.  The presence of a certain number of extinct (dormant) cometary nuclei in the NEA population is not in doubt; it is estimated that it may account for almost 6\% of the total number of known NEAs \citep{opi63,wei89,jen08,bab091}.

A cometary nucleus which was active at past and then loosed ability to generate a coma visible at any parts of its orbit is denoted by the term dormant or extinct comet \citep{wei89}. The nuclei of such comets have covered throughout evolution by the thick infusible mantle; as a consequence, a normal cometary activity was ceased \citep{whi50, whi51, opi63}. Note, extinct comet could be reactivated by the perturbing to the smaller perihelion distance or due to collision with another body \citep{wei89}. By the ground-based observations extinct cometary nuclei outwardly do not differ from asteroids. However, they may be distinguished by the dynamical properties, namely, by their orbital elements. The typical comet-like orbit of an object implies that it has a cometary origin. Finding the connection of such an object with the meteoroid stream producing active observable on the Earth meteor showers justifies the supposition that the object is an extinct comet.

Several asteroid-meteoroid complexes were identified using this “meteor sign” approach, such as the Piscid complex  \citep{bab08b}, the $iota$-Aquariid complex  \citep{bab09}, the $chi$-Scorpiid complex (possibly, the $chi$-Scorpiids relates to the $omega$-Scorpiid complex) \citep{coo73,bab13}, \, the \, \,  $sigma$-Capricornid \, \, complex\, \,  \citep{bab15b}, etc. These complexes consist of the meteoroid streams and objects of cometary origin identified among NEAs, which (these objects) may supposedly be the parent bodies of these streams. However, the parent bodies are not established for majority of known meteoroid streams. Furthermore, a search for the parents of meteoroid streams is a key stage in a recognizing of genetic associations between the minor bodies of the Solar system. We have continued the investigation of their interrelations and revealing new extinct comets among the NEAs, and present the results of a finding of related objects in the Virginid meteoroid complex. 
\section{Meteor showers of the Virginid complex}
The group of meteor showers producing by the relevant streams (sub-streams) and having the area of radiation in the Virgo constellation forms the Virginid complex. The period of activity of the nighttime showers begins in February, completely covers March-April and terminates in May, while the activity of the daytime showers falls on the period September-October. So that it is one of the largest regions of the annual meteor activity appearing, as a rule, in 165 deg. (the nighttime showers) and 345 deg.  (the daytime showers) to the West from the Sun. The region was named as the Virginid shower by \citet{hof48} and mentioned as one of the ecliptic showers visible during a year. Hoffmeister's study was the first to attempt to identify the individual streams of the Virginid region. The strongest supported shower of this complex belongs to the $theta$-Virginids and was based on 12 individual radiant determinations. It was most active on March 15 at the Solar longitude of 355.7 deg. from a radiation region of $RA$=192.4 deg. and $DEC$=-1.5 deg. \citep{hof48}. Garry Kronk in his catalogue  lists the following branches of the Virginid comlpex from February to May: $eta$-Virginids, $pi$-Virginids, $theta$-Virginigs, $alpha$-Virginids, $gamma$-Virginids and April Virginids \citep{kro88}. Some Virginid's showers were mentioned in the monographs of \citet{lov54, jen06} as well.

According the IAU MDC meteor showers database available at the website \citet{MODC} \citep[see][]{jop11,jop14,jop17,haj23} the following established showers of the Virginid complex are provided: $alpha$-Virginids (21/AVB) \citep{sou63, lin71a,lin71b, jen16}, having the activity period in 12-22 April; $h$-Virginids (343/HVI) \citep{son09,mol09,jen16,rog20},  maximum activity in 22-29 April; February $epsilon$-Virginids (506/FEV) \citep{ste13,mol13,jen16} with activity in 3-4 February and possible association with comets C/1808F1 (Pons) and C/1978T3 (Bradfield); $eta$-Virginids (11/EVI) maximal active in 18 March \citep{lin71a,jen06,jen16},  the proposed parent is comet D/1766 G1 (Helfenzrieder). Besides, some showers are given in the database which need the observational confirmation: Northern and Southern March Virginids (123/NVI, 124/SVI) \citep{sek73,kro88,kas67}, with the activity maximum in 15-19 and 17-23 March respectively; $kappa$-Virginids (509/KVI) \citep{seg13}, 27  March;  $mu$-Virginids (47/DLI) \citep{por94}, 20-29 April; Northern and Southern $gamma$-Virginids (134/NGV and 135/SGV) \citep{ter89,por94},  the maximum activity in 14 and 13 April respectively, and the possible parent bodies NEAs 2002 FC and 2003 BD44, correspondingly; etc.
 
Table~\ref{table1} lists the average values of the more important parameters of the Virginids complex. They were taken from the IAU MDC database, but only those values were selected which were obtained from the averaging of the largest number of individual radiants and orbits of the members of the given shower. Bibliographic sources are given from the IAU MDC and denoted as \citet{jen16} - J1, \citet{rog20} - R, \citet{sek73} - S2, \citet{seg13} - SE, \citet{lin71b} - L2, \citet{por94} - P, \citet{ter66,ter89} -  T, T1.

\begin{table*}
\caption{Meteor showers included in the Virginids complex. The mean geocentric and heliocentric parameters were taken from the IAU MDC database (status 2023-May).  $SunL$ is the Sun ecliptic longitude at the maximum shower activity, $RAg$ and $DECg$ are the geocentric radiant coordinates, $V_{g}$ is the geocentric velocity in [km/s], $q$ is the perihelion distance in [au], $e$ is the eccentricity, $\omega$ is the argument of perihelion, $\Omega$ is the longitude of the ascending node, $i$ is the inclination of the orbit, $N$ means the number of orbits used to find the mean shower parameters, in the last column the bibliographic source is given from the IAU MDC. The angular parameters are given in degrees in the reference system J2000.  The four streams officially named by the IAU are highlighted by bold font. Note, the showers Northern and Southern $nu$-Virginids proposed by \citet{ter66,ter89} did not included into the MDC.}
\scriptsize{
\begin{flushleft}
\begin{tabular}{lccccccccccc}
\hline
\hline
Codes&SunL&$RA_{g}$&$DEC_{g}$&$V_{g}$&$q$&$e$&$\omega$&$\Omega$&$i$&$N$&Ref.\\
\hline
{\bf 21/AVB}&32.0&203.5&2.9&18.8&0.744&0.716&247.9&30.0&7.0&12&J1\\
{\bf $\alpha$-Virds}&&&&&&&&&&&\\
\hline
{\bf 343/HVI}&38.4&203.0&-10.7&18.8&0.740&0.741&66.0&219.8&0.4&34&R\\
{\bf $h$-Virds}&&&&&&&&&&&\\
\hline
{\bf 506/FEV}&314.0&200.4&11.0&62.9&0.491&0.954&272.5&312.6&138.0&55&J1\\
{\bf Feb.$\epsilon$-Virds}&&&&&&&&&&&\\
\hline
{\bf 11/EVI}&357.0&184.8&3.9&26.6&0.460&0.812&281.0&355.7&5.4&57&J1\\
{\bf $\eta$-Virds}&&&&&&&&&&&\\
\hline
123/NVI&358.0&185.7&2.3&23.0&0.496&(0.707)&282.4&358.0&3.7&18&S2\\
Northern &&&&&&&&&&&\\
March Virds&&&&&&&&&&&\\
\hline
124/SVI&2.0&179.7&-8.5&22.9&0.565&(0.738)&81.2&182.0&6.1&13&S2\\
Southern &&&&&&&&&&&\\
March Virds&&&&&&&&&&&\\
\hline
509/KVI&6&208&-8&37.4&0.139&0.938&321&6&8&58&SE\\
$\kappa$-Virds&&&&&&&&&&&\\
\hline
49/LVI&20&210.7&-10.2&26.8&0.343&(0.870)&295.0&20.2&2.0&-&L2\\
$\lambda$-Virds&&&&&&&&&&&\\
\hline
47/DLI&39&226.8&-8.7&28.3&0.418&(0.835)&286.5&38.3&9.1&3&P\\
$\mu$-Virds&&&&&&&&&&&\\
\hline
134/NGV&24.3&180.6&17.7&11.7&0.908&(0.596)&221.7&24.3&5.2&-&T1\\
Northern&&&&&&&&&&\\
$\gamma$-Virds&&&&&&&&&&&\\
\hline
135/SGV&22.7&183.2&-15.5&13.9&0.867&(0.617)&50.0&212.7&5.0&5&P\\
Southern&&&&&&&&&&&\\
$\gamma$-Virds&&&&&&&&&&&\\
\hline
Northern&348.8&173.0&6.0&24.8&0.521&0.790&274.4
&348.8&2.3&-&T1\\
$\nu$-Virds&&&&&&&&&&&\\
\hline
Southern&340.4&170.0&4.0&28.5&0.387&0.850&109.1&160.4&0.3&-&T\\
$\nu$-Virds&&&&&&&&&&&\\
\hline
\end{tabular}
\end{flushleft}
\label{table1}}
\end{table*}

The parent comets of meteoroid streams of the Virginid complex have not been established surely. NEA 1998 SH2 was suggested as a proper parent body of the $alpha$-Virginids (21/AVB) stream \citep{jen06}. The related association of NEA 2004 CK39 and the Northern and Southern $nu$-Virginids \citep{ter66, ter89} as well of asteroid 2007 CA19 and the Northern and Southern $eta$-Virginids \citep{sek73,sek76, ter66, ter89} was found and a cometary nature of these asteroids was supposed on this base \citep{bab12, bab15c}.  In our study, we performed a search for new objects associated with the complex, among NEAs discovered till the end of 2014, as a result, four asteroids families were identified related to four sub-streams of the Virginid complex. The results are presented in this paper.

\section{Method of investigation}
We adopted a concept of formation and evolution of meteoroid streams given by \citet{bab92}. According to it, from a huge number of stream's meteoroids, only the bodies having the orbits with the heliocentric distances of ascending $R_{a}$ and descending $R_{d}$ nodes close to 1 AU can intersect the Earth's orbit. Alongside with this, for the majority of NEAs this condition realizes four times over one Kozai-Lidov cycle of the orbital argument of perihelion. If an asteroid is actually a non-active comet, then in the past it may have created a meteoroid stream. Theoretically, such a stream may produce four meteor showers observable on the Earth -  the nighttime showers with northern and southern branches and the daytime showers also with northern and southern branches. Having the orbital elements for which the nodal distances $R_{a}$ or $R_{d}$  are close to 1 AU, one can calculate the theoretical radiants of the possible meteor showers. The identification of theoretical showers with observable showers, fireballs/meteors and other objects most likely confirms the relationship between these showers and the studied asteroid as well as the cometary nature of the asteroid.
The orbits of theoretical meteor showers can be obtained by calculation of the orbital evolution of the proposed parent body – near-Earth asteroid. The orbital evolution is calculated using various methods of numerical integration of the motion equation for the time equal to one cycle of variation of the argument of perihelion of the asteroid orbit. In the case of NEAs, this time interval is usually between 5 and 12 thousand years. The methods of \citet{eve74} and Halphen-Goryachev \citep{gor37} are the most using for the orbital evolution calculation. 
\section{The near-Earth asteroids associated with Virginids complex}
Only asteroids moving on the comet-like orbits and crossing the Earth's orbit are subject of the investigation. The Tisserand parameter $T_{j}$ is using for classification of orbits \citep{kre69,kre82,koz92}. While the value of the Tisserand parameter satisfies the condition $T_{j}\leq$3.12, then the orbit is classified as comet-like, when $T_{j}>$3.12 the orbit relates to asteroidal type \citep{kre69,jew12}.
Applying Tisserand condition $T_{j}\leq$3.12,  29 NEAs from  JPL's Solar System Dynamics database \citep{nasa} was selected and their motion was calculated by RADAU19 integrator \citep{eve74}. Gravitational perturbations of major planets Mercury-Jupiter were taken into account as well; the relevant planetary data were taken \, \,  from \, \,  the \, \, JPL \, \, DE403 \, \, ephemeris \citep{sta95,nasa}. 

\subsection{Investigation of orbital evolution}

\begin{figure}
\begin{center}
\includegraphics*[width=0.95\textwidth]{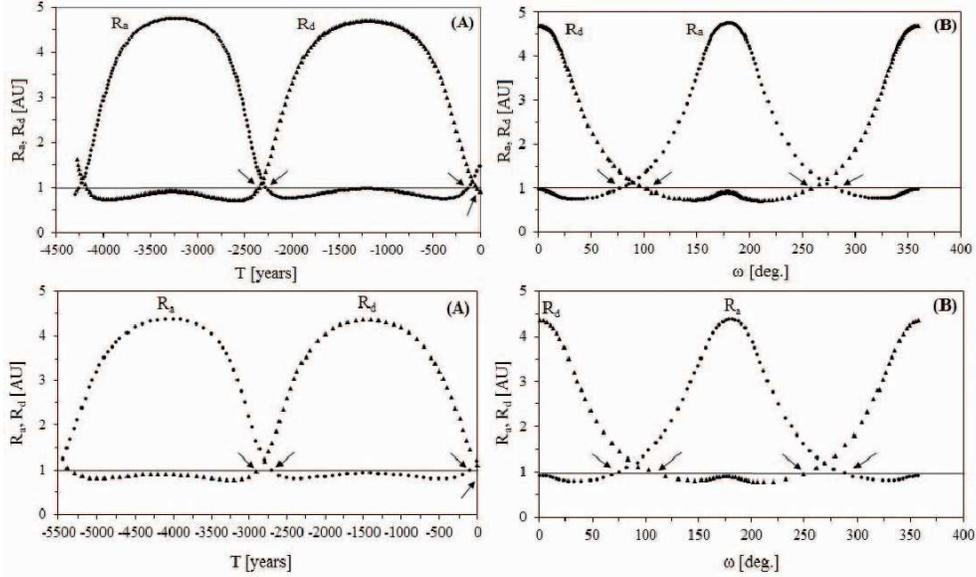}
\end{center}
\caption{ Variations of nodal distances $R_{a}$ and $R_{d}$ of the orbit of asteroids associated with the $alpha$-Virginids meteoroid stream. Top, the NEA 2010 EK44, bottom the  NEA 2014EQ.  The points corresponding to the conditions $R_{a}$ =$R_{d}$ =1 AU, i.e. intersection of the Earth's orbit and the asteroid orbit, are indicated by arrows. }
\label{fig1}
\end{figure}

\begin{figure}
\begin{center}
\includegraphics[width=0.95\textwidth]{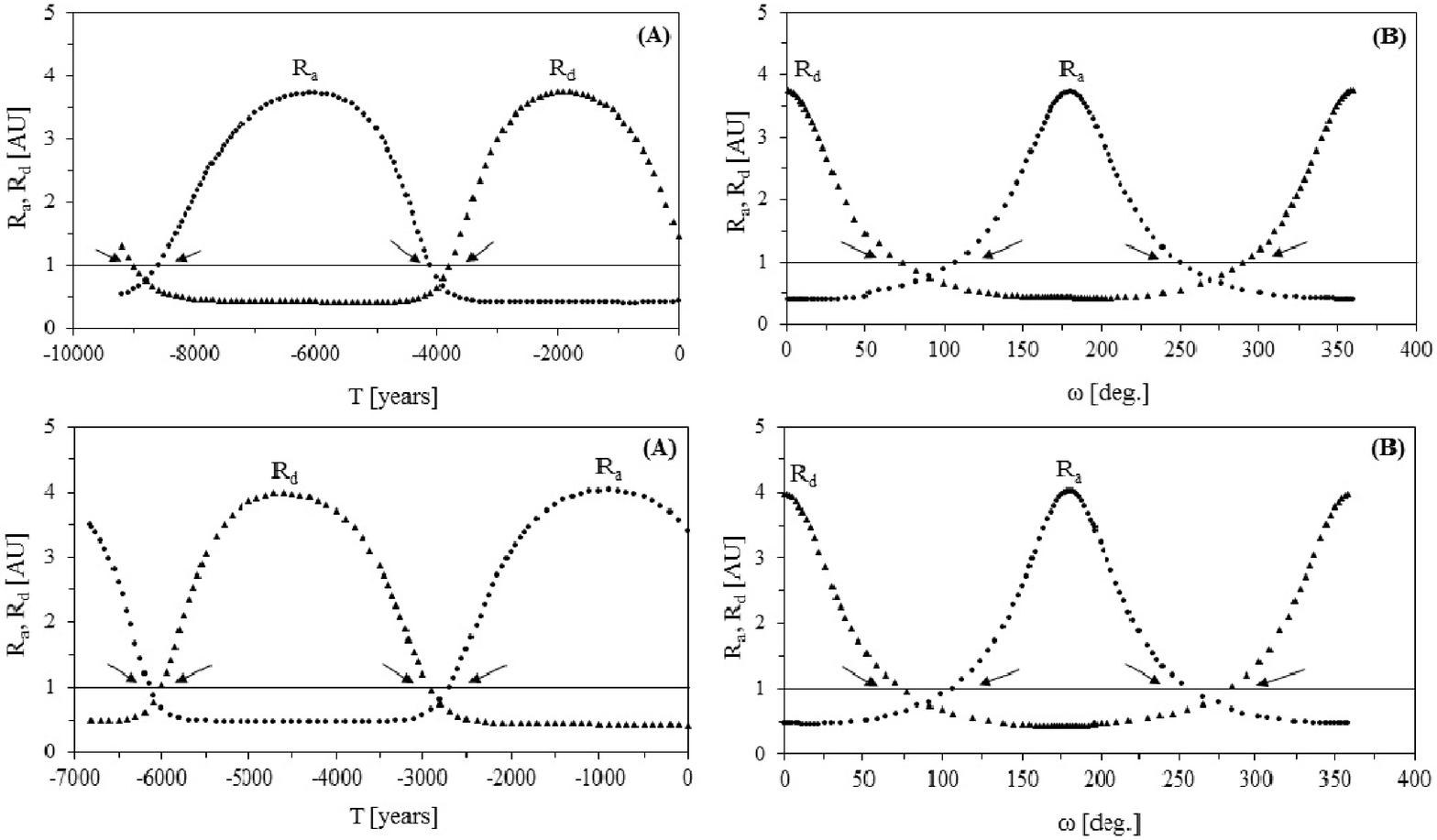}
\end{center}
\caption{Variations of nodal distances $R_{a}$ and $R_{d}$ of the orbit of asteroids associated with the $eta$-Virginids meteoroid stream. Top the NEA 2010 FJ81, bottom the NEA 2014 VC10.  The points corresponding to the conditions $R_{a}$ =$R_{d}$ =1 AU, i.e. intersection of the Earth's orbit and the asteroid orbit, are indicated by arrows.  }
\label{fig2}
\end{figure}

\begin{figure}
\begin{center}
\includegraphics[width=0.95\textwidth]{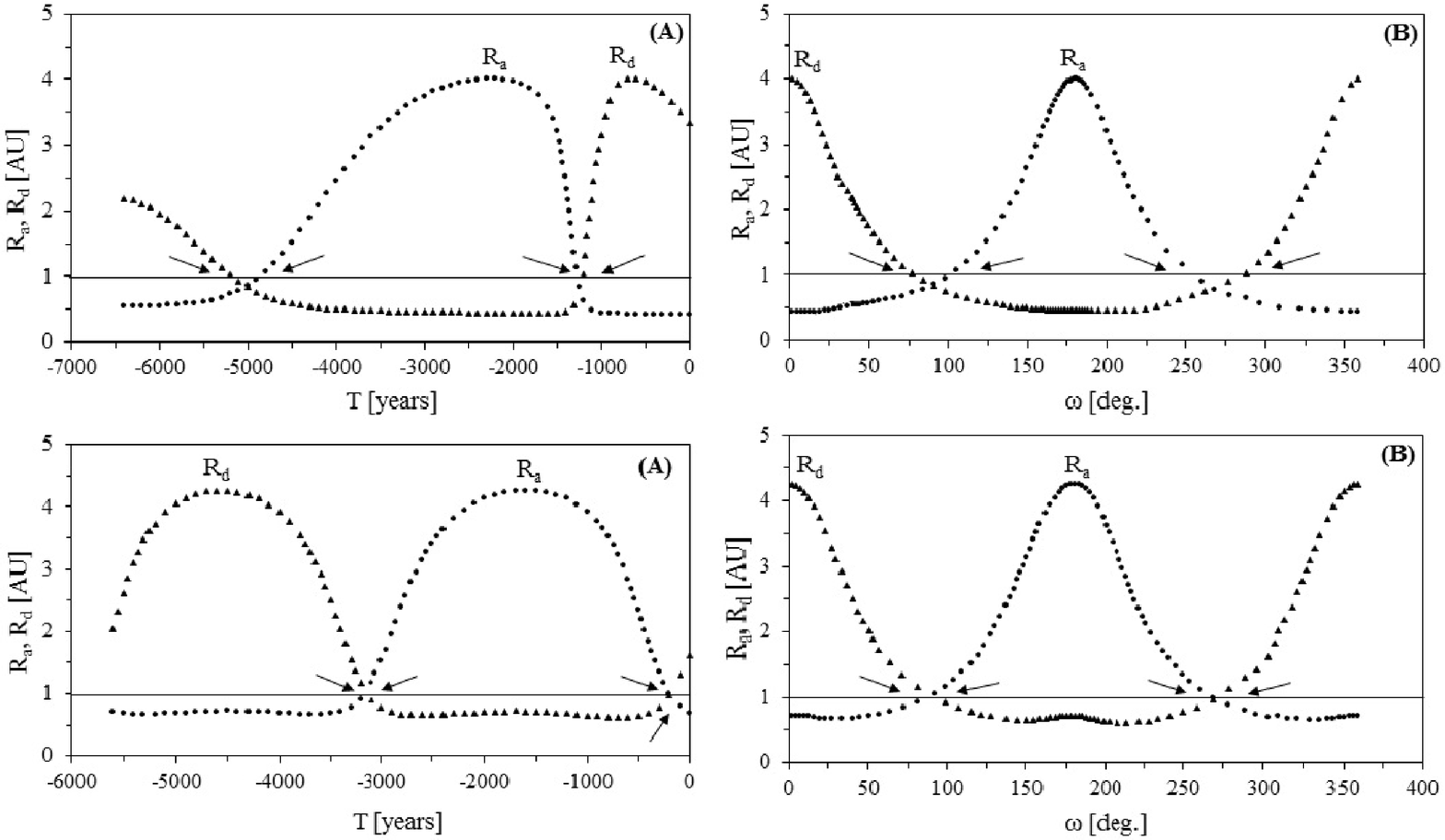}
\end{center}
\caption{ Variations of nodal distances $R_{a}$ and $R_{d}$ of the orbit of asteroids associated with the $nu$-Virginids meteoroid stream. Top the NEA 2013 TR13, bottom the NEA 2013 CU82. The points corresponding to the conditions $R_{a}$ =$R_{d}$ =1 AU, i.e. intersection of the Earth's orbit and the asteroid orbit, are indicated by arrows. }
\label{fig3}
\end{figure}

\begin{figure}
\begin{center}
\includegraphics[width=0.95\textwidth]{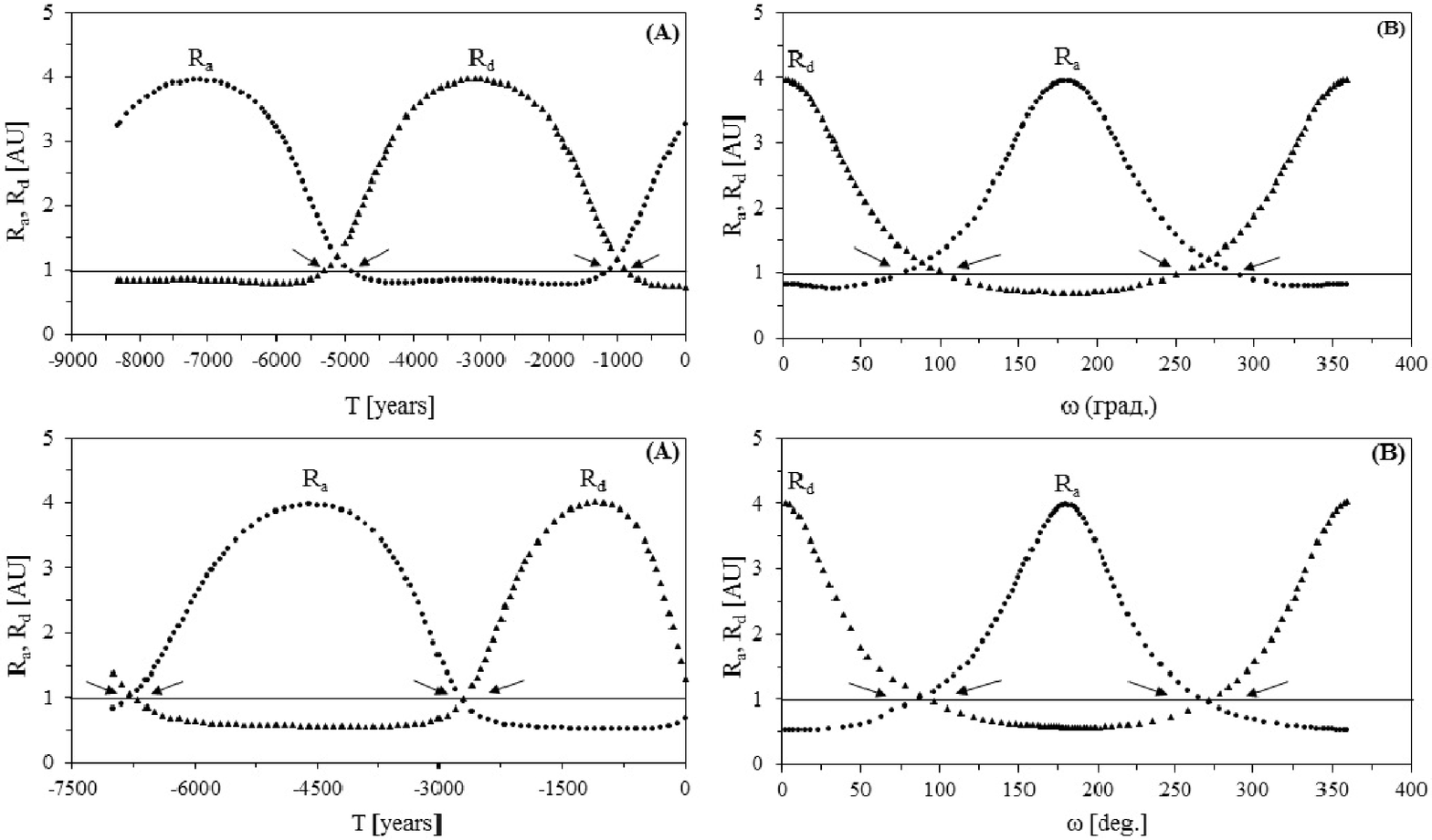}
\end{center}
\caption{ Variations of nodal distances $R_{a}$ and $R_{d}$ of the orbit of asteroids associated with the $mu$-Virginids meteoroid stream. Top the NEA 2010 GO33, bottom the NEA 2011 VG9. The points corresponding to the conditions $R_{a}$ =$R_{d}$ =1 AU, i.e. intersection of the Earth's orbit and the asteroid orbit, are indicated by arrows.  }
\label{fig4}
\end{figure}

As a result of calculation of orbital evolution it was found that 29 asteroids listed in Table~\ref{table2}, cross the Earth’s orbit for four times during one cycle of $\omega$ variation. It means that the sizes of $R_{a}$ and $R_{d}$ of asteroids orbits have the values close to 1 AU four times – two cases at each of the nodes. The variability of the length $R_{a}$ and $R_{d}$ depending on time and argument of perihelion is illustrated in figures below. These graphs are similar for all considered NEAs so the plots for a pair of selected asteroids are presented. For asteroids 2010 EK44 and 2014 EQ associated with the $alpha$-Virginid meteoroid stream it is shown in Figs. ~\ref{fig1}; for asteroids 2010 FJ81 and 2014 VC10 associated with the $eta$-Virginid meteoroid stream it is shown in Figs.~\ref{fig2}; for asteroids 2013 TR13 and 2013 CU82 associated with the $nu$-Virginid meteoroid stream it is shown in Figs. ~\ref{fig3}; and for pair asteroids 2010 GO33 and 2011 VG9 associated with the $mu$-Virginid meteoroid stream it is shown in Figs.~\ref{fig4}.


A set of the orbital elements which corresponds to the Earth’s crossing positions found from the calculated orbital evolution serves as the initial data for calculation of the coordinates of theoretical geocentric radiants (right ascension $\alpha_{g}$ and declination $\delta_{g}$), the geocentric velocities $V_{g}$, the Solar longitudes $L_{\odot}$ and relevant dates of the activity of theoretical showers.  

\subsection{General results}

A computerized search for similar theoretically predicted and observed showers was carried out in the published meteor/fireball catalogues:
\citet{ter66, ter89} (T, T1), \citet{lin71a,lin71b} (L1, L2), \citet{leb73} (L), \citet{kas67} (K), \citet{jen16} (J1),
\citet{rud15} (EDM1), \citet{rud14} (EDM2), \citet{sou63} (SH), \citet{son09} (SON),  \citet{sek73,sek76} (S1, S2), \citet{gaj04} (G),
\citet{coo73} (C), \citet{por94} (P), etc.  Up-present orbital elements of 29 considered NEAs
(Table~\ref{table2}) were also included into database where we performed a search. Additionally, we performed this search among databases of individual meteors/fireballs and known meteorites: \citet{MODC} (MODC), \citet{hal96}  (MORP), \citet{mcc78} (PN), EN – fireballs detected by the European fireball network, \citet{cep61}  (C1), \citet {spu03} (SP).

Under a comparison of theoretical and observable parameters we required that the difference in radiants not exceed  $\pm$10 deg. in both right ascension and declination, the difference in geocentric velocities should be $\Delta V_{g}\pm$5 km/s, and the periods of activity may differ no more than $\pm$15 days   \citep[see e.g.][]{bab09}.

A closeness of the theoretical and observable orbits was examined using the \citet{sou63} $D$-function, $D_{SH}$, being a measure of the similarity between two orbits. It is accepted to consider two orbits similar when the condition $D_{SH}\leq$ 0.25 is fulfilled \citep{sou63}.
  
As a result of a search it turned out that considered NEAs form four groups associated with four sub-streams of the Virginid complex, asteroids parameters are given in the separate sections of Table 2. In addition to the elements of Keplerian orbits we provide the absolute magnitudes, equivalent diameters of the NEAs as well as the number of intersections of the asteroid's orbit with the Earth's orbit  during one Kozai-Lidov cycle. 

The diameters denoted by asterisk in Table 2 are accessible at the published JPL SSD database. At the cases of a lack of size estimation the diameters were calculated by the following empirical equation \citet{har02}
\begin{equation}
d=\frac{1329}{\sqrt{p}\cdot 10^{0.2H}},
\end{equation}
where $p$ is the geometrical albedo of an asteroid. Very dark asteroids of $C, P$, and $D$ types having a low albedo within 0.02-0.12 very likely might be extinct comets \citep{jew92}. The diameters of asteroids were estimated using the average value of albedo $p$=0.07.

\begin{table*}
\caption{ The basic features of NEAs associated with the Virginid complex. In the first column  the stream or NEA designations are given, the next ones contain Keplerian orbital elements (for NEAs MJD equal to 2014.12.09), $H$ is the absolute magnitude, $d$ is an equivalent diameter, $T_{j}$  is Tisserend parameter, $Ni$  is the number of intersections of the stream’s and asteroid’s orbits with the Earth’s orbit during one Kozai-Lidov cycle.  All angular elements of the meteoroid streams and NEAs are given in the J2000 reference system.}
\scriptsize{
\begin{flushleft}
\begin{tabular}{lccccccccccc}
\hline
\hline
Object&$a$,&$e$&$q$,&$i^{\circ}$&$\omega^{\circ}$&$\Omega^{\circ}$&$\pi^{\circ}$&$H$&	$d$,&$Ni$&$T_{j}$\\
&AU&&AU&&&&&&km&&\\
\hline
21/AVB &2.550&0.716&0.744&	7.0&247.9&30.0&277.9&-&-&4&3.01\\
2009 HS44&2.572&0.702&0.766&2.4&209.3&73.1&282.4&26.6&0.02&4&3.02\\
2009 SM98&2.432&0.700&0.730&7.5&166.9&110.9&277.8&22.3&0.17&4&3.11\\
2010 EK44&2.835&0.772&0.646&11.2&180.8&107.6&288.4&19.8&0.55&4&2.76\\
(PHA)&&&&&&&&&&&\\
2010 FD7&2.584&0.685&0.813&6.2&357.5&269.2&266.8&22.1&0.08*&4&3.04\\
2011 BE38&2.629&0.711&0.759&8.2&321.4&302.5&264.0&18.4&1.07&4&2.97\\
(PHA) &&&&&&&&&&&\\
2011 EF17&2.345&0.743&0.603&4.2&4.0&282.3&286.3&19.2&0.73&4&3.12\\
(PHA) &&&&&&&&&&&\\
2012 SW20&2.460&0.680&0.788&10.2&209.8&62.1&272.0&19.7&0.59&4&3.11\\
(PHA) &&&&&&&&&&&\\
2012 TT256&2.666&0.729&0.724&11.1&106.3&172.0&278.2&19.5&0.63&4&2.91\\
2013 TE135&2.361&0.759&0.569&6.0&182.4&102.9&285.4&19.9&0.52&4&3.08\\
(PHA) &&&&&&&&&&&\\
2014 EQ12&2.637&0.744&0.675&11.0&178.0&92.3&270.3&21.1&0.31&4&2.91\\
(PHA) &&&&&&&&&&&\\
2014 GN1&2.649&0.749&0.665&3.0&193.1&80.3&273.4&24.6&0.06&4&2.91\\
1998S H2&2.693&0.723&0.760&2.5&13.3&261.0&274.3&20.8&0.35&4&2.91\\
(PHA) &&&&&&&&&&&\\
1998 FR11&2.791&0.713&0.802&6.6&130.3&157.9&288.2&16.5&2.52&4&2.91\\
1999 UZ5&2.563&0.799&0.532&10.3&226.6&66.7&293.3&21.9&0.21&4&2.85\\
\hline
\hline
11/EVI &2.470&0.812&0.460&5.4&281.0&355.7&	276.7&-&-&4&2.91\\
2010 FJ81&3.598&0.684&1.138&42.5&97.4&171.7&269.1&20.5&0.42*&4&2.34\\
2012 BJ134&2.258&0.833&0.377&21.2&144.6&138.2&282.8&18.2&1.15&4&2.99\\
2014 VC10&2.231&0.817&0.408&11.8&84.0&196.7&280.7&19.4&0.67&4&3.07\\
2003 FB5&2.512&0.788&0.532&5.3&358.4&288.4&286.8&23.4&0.10&4&2.92\\
1995 EK1&2.266&0.776&0.508&8.8&355.5&296.7&292.2&17.9&1.34&4&3.12\\
2007 CA19&2.787&0.827&0.483&9.6&174.8&97.6&272.4&17.6&0.90&4&2.80\\
(PHA) &&&&&&&&&&&\\
\hline
\hline
$\nu$-Virds&2.342&0.822&0.394&5.8&289.6&336.6&271.1&-&-&4&2.99\\
2013 TR135&2.224&0.809&0.424&1.8&243.4&18.0&261.4&23.3&0.11&4&3.11\\
2013 CU82&2.493&0.774&0.562&10.8&317.8&303.3&261.0&21.1&0.31&4&2.95\\
2002 BK25&2.296&0.749&0.576&11.9&156.5&103.6&260.1&18.1&1.22&4&3.12\\
(PHA) &&&&&&&&&&&\\
2002 TZ59&2.597&0.842&0.411&16.5&7.4&249.7&257.1&22.9&0.13&4&2.79\\
2004 CK39&2.298&0.829&0.392&12.3&359.5&265.1&264.6&19.3&0.69&4&2.99\\
\hline
\hline
47/DLI &2.531&0.835&0.418&9.1&286.5&39.3&325.8&	-&-&4&2.81\\
2010 XD11&2.210&0.854&0.322&16.4&118.0&196.1&314.2&18.09&1.21&4&3.01\\
2010 GO33&2.409&0.699&0.724&19.1&143.6&153.6&297.2&19.19&0.31*&4&3.08\\
2011 VG9&2.275&0.776&0.509&1.3&236.1&65.9&302.0&22.18&0.18&4&3.11\\
2003 WW26&2.388&0.798&0.482&6.3&57.7&254.6&312.3&22.21&0.20&4&2.99\\
\hline
\end{tabular}
\end{flushleft}
\label{table2}}
\end{table*}

In Table 2, the common features of 29 NEAs (including two asteroids 2007CA19 and 2004 CK39 previously established by \citet{bab12,bab15c}) are the cometary nature of their orbits and close values of the longitudes of perihelion within each group. Furthermore, 9 asteroids denoted as PHA are classified as potentially hazardous ones \citep{nasa}.
 
\section{Identification of the theoretically predicted and observable showers}

\subsection{Asteroids associated with the $alpha$-Virginids stream}

As a result of a search it was found that a set of 14 NEAS (first section of Table 1) have close association with the meteoroid stream generating the nighttime $alpha$-Virginids (21/AVB) shower (the Northern branch), and the $h$-Virginids (343/HVI) and $chi$-Virginids (the Southern branch). The parameters of the HVI \citep{jen16,son09} and the $chi$-Virginids \citep{ter89} showers are also matching with the account of the daily radiant drift $dRA$=0.11, $dDEC$=-0.27 \citep{son09}; therefore it may be assumed that it is one shower. Besides, the related association of NEAs: 1998 SH2, 2009 HS44, 2012 SW20, 2014 GN1, 1999 UZ5, 2009 SM98, 2014 EQ12, 2010 EK44, 2013 TE135, 1998 FR11, 2010 FD7, 2011 EF17, 1995 EK1, 2003 FB5 and 2011 BE38 with the theoretical and observable showers is found on the base of relevant radiants and the present-date orbits of asteroids. A proposed link of the stream with asteroid 1998 SH2 is also confirmed. 

The results of a search for asteroids 2010 EK44 and 2010 FD7 are presented in Tables~\ref{table3} and ~\ref{table4}. For the rest asteroids of this set the outcomes are identical. The orbital elements, radiants, Solar longitudes and activity dates of theoretically predicted showers are given by bold font in Tables 3 and 4; the Northern branch of nighttime shower is denoted as NNS, the Southern branch of nighttime shower is denoted as SNS, the Northern branch of daytime shower is denoted as NDS, and the Southern branch of daytime shower is denoted as SDS. Catalogues and papers where active observable showers and fireballs, known meteorites  identical with the theoretical showers were found listed in Tables 3 and 4 as T1 \citep{ter89}, L1, L2 \citep{lin71a,lin71b}, L \citep{leb73}, K \citep{kas67}, J1 \citep{jen16}, SH \citep{sou63}, SON \citep{son09}, C1 \citep{cep61}, SP \citep{spu03}, MORP \citep{hal96}, PN \citep{mcc78}. The present-date orbits of the asteroids are taken  in \citet{nasa} denoted in Tables 3 and 4 as JPL, their  radiants calculated in this work denoted as tw.
 
It turned out that features of some studied NEAs are close to  features of theoretical showers. The NEAs present-day orbital elements were given from JPL SSD database \citep{nasa}, their geocentric radiants, velocities and Solar longitudes were calculated by us using the orbital elements. A closeness of present-day orbits (and radiants) of some NEAs to theoretical orbits (and radiants) points to a genetic relationship between asteroid as well.  As seen, the values of the $D_{SH}$ criterion confirm a similarity of the theoretical and observable orbits. Identification of the theoretical showers with the observable objects is approving by a closeness of the radiant positions, velocities and activity dates. It allows concluding that the association of the asteroids with these showers exists and, consequently, NEAs under consideration very likely have in fact a cometary origin. Moreover, these objects may be considered as the parent bodies of the stream or they can be extinct fragments of a larger comet-progenitor of the stream. Thus this meteoroid stream contains large-sized remnants of the parent comet which presently are in dormant stage. 

\begin{table*}
\caption{ Meteor showers, fireballs and NEAs associated with NEA 2010 EK44 (J2000.0).}
\scriptsize{
\begin{flushleft}
\begin{tabular}{lcccccccccccl}
\hline
Objects&$q$&$e$&$i^{\circ}$&$\Omega^{\circ}$&$\omega^{\circ}$&$L_{\odot}^{\circ}$&Date&$\alpha_{g}^{\circ}$&$\delta_{g}^{\circ}$&$V_{g}$&$D$&Ref.\\	
&AU&&&&&&&&&km/s&$_{SH}$&\\								
\hline							
{\bf NNS}&{\bf 0.648}&{\bf 0.771}&{\bf 9.3}&{\bf 29.2}&{\bf 259.2}&{\bf 29.2}&{\bf 19Apr}&{\bf 207.7}&{\bf 2.9}&{\bf 22.2}&-&{\bf tw}\\
21/AVB&0.724&0.716&7.0&30.0&247.9&30.0&20Apr&203.5&2.9&18.8&0.17&J1\\
As.8&0.657&0.700&11.0&28.0&260.0&28.0&18Apr&208.0&6.0&21.2&0.08&K\\
299&0.609&0.697&3.7&16.1&267.1&16.1&06Apr&195.7&-0.8&21.0&0.19&MORP\\
300&0.715&0.680&2.6&18.0&252.9&18.0&07Apr&190.2&0.6&18 .5&0.13&MORP\\
080477&0.564&0.770&10.6&18.3&269.5&18.3&08Apr&203.0&4.74&26.4&0.09&EN\\
1998SH2&0.746&0.723&2.5&13.3&261.0&13.3&03Apr&184.6&2.8&19.3&0.20&JPL,tw\\
\hline
{\bf SNS}&{\bf 0.646}&{\bf 0.772}&{\bf 10.7}&{\bf 209.2}&{\bf 79.2}&{\bf 29.2}&{\bf 19Apr}&{\bf 197.3}&{\bf -23.8}&{\bf 22.4}&-&{\bf tw}\\
343/HVI&0.742&0.734&0.8&219.0&67.64&39.0&29Apr&204.2&-11.6&18.7&0.20&SON\\
1999UZ5&0.515&0.799&10.3&226.6&66.7&46.6&07May&-&-&-&0.16&JPL\\
2009HS44&0.766&0.702&2.4&209.2&73.1&29.2&19Apr&193.8&-10.8&18.2&0.20&JPL,tw\\
2012SW20&0.786&0.679&10.2&209.8&62.1&29.8&19Apr&186.7&-24.7&17.6&0.20&JPL,tw\\
2014GN1&0.665&0.748&3.0&193.1&80.3&13.1&02Apr&184.9&-7.0&21.1&0.20&JPL,tw\\
\hline
{\bf NDS}&{\bf 0.645}&{\bf 0.772}&{\bf 10.9}&{\bf 187.8}&{\bf 100.6}&{\bf 187.8}&{\bf 01Oct}&{\bf 196.9}&{\bf 9.6}&{\bf 22.5}&-&{\bf tw}\\ 
2010EK44&0.645&0.772&11.2&180.8&107.6&180.8&23Sep&191.8&12.7&22.0&0.02&JPL,tw\\
2013TE135&0.569&0.759&6.0&182.4&102.9&182.4&25Sep&185.5&6.1&23.0&0.12&JPL,tw\\
2014EQ12&0.676&0.743&11.0&178.0&92.3&178.0&21Sep&189.6&14.1&21.5&0.20&JPL,tw\\
\hline
{\bf SDS}&{\bf 0.651}&{\bf 0.770}&{\bf 9.6}&{\bf 7.0}&{\bf 281.4}&{\bf 187.0}&{\bf 30Sep}&{\bf 185.1}&{\bf -17.3}&{\bf 22.2}&-&{\bf tw}\\
1995EK1&0.508&0.776&8.8&355.5&296.7&-&-&-&-&-&0.19&JPL\\
2010FD7&0.813&0.685&6.2&357.5&296.2&177.5&20Sep&186.7&-16.4&17.2&0.19&JPL,tw\\
2011EF17&0.604&0.742&4.2&4.0&282.3&184.0&27Sep&183.7&-8.1&21.7&0.11&JPL,tw\\
\hline
\end{tabular}
\end{flushleft}
\label{table3}}
\end{table*}

\begin{table*}
\caption{ Meteor showers, fireballs, meteorites, and NEAs associated with NEA 2010 FD7 (J2000.0).}
\scriptsize{
\begin{flushleft}
\begin{tabular}{lcccccccccccl}
\hline
Objects&$q$&$e$&$i^{\circ}$&$\Omega^{\circ}$&$\omega^{\circ}$&$L_{\odot}^{\circ}$&Date&$\alpha_{g}^{\circ}$&$\delta_{g}^{\circ}$&$V_{g}$&$D$&Ref.\\ 
&AU&&&&&&&&&km/s&$_{SH}$&\\
\hline 
{\bf NNS}&{\bf 0.819}&{\bf 0.683}&{\bf 7.2}&{\bf 30.3}&{\bf 236.4}&{\bf 30.3}&{\bf 20Apr}&{\bf 197.2}&{\bf 9.5}&{\bf 16.3}&-&{\bf tw}\\
21/AVB&0.691&0.732&4.2&21.7&240.4&21.7&11Apr&185.4&9.6&16.8&0.20&SH\\
21/AVB&0.750&0.681&0.7&28.9&247.5&28.9&19Apr&195.6&-5.3&16.6&0.11&L2\\
21/AVB&0.809&0.684&3.1&31.8&239.7&31.8&22Apr&195.8&0.4&16.4&0.10&L1\\
21/AVB&0.724&0.716&7.0&30.0&247.9&30.0&20Apr&203.5&2.9&18.8&0.07&J1\\
300&0.715&0.680&2.6&18.0&252.9&18.0&07Apr&190.2&0.6&18.5&0.13&MORP\\
650404&0.777&0.650&0.7&16.0&244.0&16.0&05Apr&183.0&0.5&16.2&0.15&PN\\
Pribram&0.790&0.674&10.4&17.8&241.6&17.8&07Apr&192.1&17.4&17.5&0.11&P1\\
Neusch-&0.792&0.670&11.4&17.5&241.2&17.5&06Apr&192.3&19.5&17.5&0.12&SP\\
wanstein&&&&&&&&&&&&\\
1998SH2&0.746&0.723&2.5&13.3&261.0&13.3&03Apr&184.6&2.8&19.3&0.15&JPL,tw\\
\hline
{\bf SNS}&{\bf 0.881}&{\bf 0.659}&{\bf 8.9}&{\bf 221.1}&{\bf 45.7}&{\bf 41.1}&{\bf 1May}&{\bf 184.7}&{\bf -26.5}&{\bf 14.5}&-&{\bf tw}\\
$\chi$-Virds&0.718&0.725&1.4&202.9&70.7&22.9&12Apr&190.0&-7.0&19.1&0.12&T1\\ 
670426&0.829&0.710&6.2&217.0&56.4&37.0&27Apr&191.0&-19.3&16.5&0.11&PN\\
2009HS44&0.766&0.702&2.4&209.2&73.1&29.2&19Apr&193.8&-10.8&18.2&0.22&JPL,tw\\
2012SW20&0.786&0.679&10.2&209.8&62.1&29.8&19Apr&186.7&-24.7&17.6&0.12&JPL,tw\\
\hline
{\bf NDS}&{\bf 0.866}&{\bf 0.665}&{\bf 9.2}&{\bf 134.4}&{\bf 132.4}&{\bf 134.4}&{\bf 07Aug}&{\bf 169.6}&{\bf 28.8}&{\bf 14.8}&-&{\bf tw}\\
1998FR11&0.801&0.713&6.6&130.3&157.9&130.3&02Aug&184.3&20.5&11.4&0.22&JPL,tw\\
\hline
{\bf SDS}&{\bf 0.891}&{\bf 0.655}&{\bf 6.1}&{\bf 309.9}&{\bf 316.9}&{\bf 129.9}&{\bf 02Aug}&{\bf 154.5}&{\bf -7.1}&{\bf 13.4}&-&{\bf tw}\\
2011BE38&0.757&0.711&8.2&321.4&302.5&141.4&14Aug&152.3&-5.5&17.5&0.15&JPL,tw\\
\hline
\end{tabular}
\end{flushleft}
\label{table4}}
\end{table*}

A search has also shown a close association of these asteroids with the theoretical and observable showers, as well their mutual link. The orbit of PHA 1998 SH2 at present is matching to the northern nighttime shower 21/AVB. The orbits of asteroids 200 9HS44, 2012 SW20, 2014 GN1, and 1999 UZ5 correspond to the southern nighttime shower 343/HVI (and $chi$-Virginids). Regarding the theoretical daytime showers, it was defined that the present-day orbits of  NEAs 2009 SM98, 2014 EQ12, 2010 EK44, 2013 TE135, and 1998 FR11 are consistent with the northern branch of the daytime shower,  while the orbits of 2010 FD7, 2011 EF17, 1995 EK1, 200 3FB5, and 2011 BE38 – with the southern branch of the daytime shower. Consequently, possible incomings of asteroids into the Earth's atmosphere will occur at the period of activity of the relevant shower and will have characteristics close to the features of this shower. For instance, it is theoretically predicted that the date of presumed impact of the potentially hazardous object 2010 EK44 will be 23 September, the geocentric velocity will be 22 km/s, and the equatorial coordinates of the point from which it will be directed to the Earth will be  $\alpha_{g}$=191.8 deg.  and $\delta_{g}$=12.7 deg.

\noindent{\it 5.1.1. A relationship with the meteorites}

Notably that the orbits, radiants, velocities and the fall dates of Pribram and  Neuschwanstein meteorites are similar with the northern nighttime showers of asteroids 2010 FD7 (Table 4), 2011 BE38, 2012 SW20, and 2014 EQ12. Pribram meteorite fall in 7 April 1959 on the Czech territory  \citep{cep61} and Neuschwanstein meteorite dropped in 6 April 2002 in the territory of the Germany-Austria \citep{spu03}. The heliocentric orbits of the meteorites are identical so it was supposed that the stream of meteorite objects having the Earth's crossing orbit was found \citep{spu03}. However, later \citet{obe03} have shown that an assumption on a common parent body of these meteorites is unlikely because of different classification of the meteorites types. Furthermore, the different duration of their cosmic rays radiation was established, for Pribram it is 19 million years \citep{obe03}. Since cometary meteoroid streams have significantly less age then these meteorite objects very likely separated from the parents more early than the stream was formed. Consequently, they very probably belong to the one stream of objects however were produced by the different parent bodies. Under this, they were detached from the parents more early then usually the meteoroids of cometary streams are ejected from the parent comets \citep{spu03}.

Found association of these meteorites with the meteor showers related to considered asteroids points to that the supposed stream of meteorites objects matches with the alpha-Virginids meteoroid stream. Alongside with a huge number of small meteoroids the asteroids of a cometary origin, i.e. extinct cometary nuclei, are moving within the stream and all they were generated by the one common parent body. Besides, the objects of an asteroidal origin of a larger age’s presence in the stream, under this, their parents differ from both the parent of the cometary objects of the stream and between themselves.

\subsection{ Asteroids associated with the $eta$-Virginid stream}

As a result of a search for the observable showers (fireballs, meteors), and NEAs close to theoretically predicted showers it was found that 6 NEAs listed in the second section of Table 1 coincide with the meteoroid stream producing the nighttime Northern and Southern  $eta$-Virginids (11/EVI) \citep{sek73,sek76, ter66,ter89}. Association of the potentially hazardous asteroid 2007 CA19 included to this group with the Northern and Southern $eta$ -Virginids was previously found by \citet{bab15c}. The Northern and Southern branches of the theoretical daytime shower were identified with 29 and 22 meteors respectively from the MODC database. Coming from their mean radiants positions and the activity dates it was proposed to design they as Daytime October $eta$- and $beta$-Virginids.  Besides, Southern branch of the theoretical daytime shower was identified with the observable Daytime $psi$-Virginids shower (240/DFV) \citep{sek76}. These observable daytime showers were identified with the theoretical daytime showers also of 6 asteroids of this sample.
 
The results of identification of theoretically predicted showers with the observable showers, fireballs/meteors, and NEAs for asteroids 2014 VC10 and 2010 FJ81 are presented in Tables~\ref{table5} and ~\ref{table6}. For the rest asteroids of this group the outcomes are identical. All designations in Tables 5 and 6 are the same as in Tables 3 and 4. 

\begin{table*}
\caption{Meteor showers, fireballs/meteors, and NEAs associated with NEA 2014 VC10 (J2000.0).}
\scriptsize{
\begin{flushleft}
\begin{tabular}{lcccccccccccl}
\hline
Objects&$q$&$e$&$i^{\circ}$&$\Omega^{\circ}$&$\omega^{\circ}$&$L_{\odot}^{\circ}$&Date&$\alpha_{g}^{\circ}$&$\delta_{g}^{\circ}$&$V_{g}$&$D$&Ref.\\ 
&AU&&&&&&&&&km/s&$_{SH}$&\\
\hline
{\bf NNS}&{\bf 0.458}&{\bf 0.795}&{\bf 3.3}&{\bf 357.9}&{\bf 282.9}&{\bf 357.9}&{\bf 18Mar}&{\bf 185.1}&{\bf 1.5}&{\bf 26.2}&-&{\bf tw}\\
11/EVI&0.464&0.812&5.4&355.7&281.0&355.7&16Mar&184.8&3.9&26.6&0.07&J1\\
Virds&0.365&0.914&3.8&348.5&286.5&348.5&09Mar&178.1&4.4&31.0&0.17&G\\
N.$\eta$-Virds&0.495&0.707&3.7&357.3&282.4&357.3&18Mar&185.1&2.3&23.0&0.10&S2\\
N.$\eta$-Virds&0.377&0.831&3.9&352.6&291.5&352.6&13Mar&184.0&2.0&28.7&0.10&T1\\
Virds&0.349&0.820&6.0&356.0&297.0&356.0&16Mar&188.0&1.0&27.8&0.20&K\\
As.112&0.433&0.870&6.1&355.5&285.6&355.5&16Mar&185.3&2.6&29.6&0.09&L\\
140378&0.390&0.870&1.3&352.8&287.9&352.8&13Mar&182.0&0.53&29.7&0.11&EN\\
710319&0.537&0.710&5.6&358.0&276.9&358.0&18Mar&184.0&5.7&22.6&0.14&PN\\
680306&0.389&0.860&3.7&345.0&287.4&345.0&05Mar&176.0&5.3&29.3&0.15&PN\\
1995EK1&0.508&0.776&8.8&355.5&296.7&-&-&-&-&-&0.19&JPL\\
2003FB5&0.533&0.788&5.3&358.4&288.4&-&-&-&-&-&0.12&JPL\\
\hline
{\bf SNS}&{\bf 0.427}&{\bf 0.809}&{\bf 5.0}&{\bf 174.4}&{\bf 106.4}&{\bf 354.4}&{\bf 15Mar}&{\bf 180.1}&{\bf -5.3}&{\bf 27.2}&-&{\bf tw}\\
S.$\eta$-Virds&0.566&0.738&6.1&182.0&91.2&2.0&22Mar&179.6&-8.0&22.9&0.19&S2\\
S. $\eta$-Virds&0.499&0.706&2.6&171.1&101.8&351.1&11Mar&176.1&-1.6&22.9&0.17&S3\\
S.$\eta$-Virds&0.396&0.904&12.3&174.5&105.4&354.5&15Mar&178.6&-10.6&31.2&0.16&T\\
As.2&0.455&0.82&6.0&175.0&100.0&355.0&15Mar&177.0&5.0&26.7&0.09&K\\
790303&0.366&0.865&0.2&162.8&111.1&342.8&3Mar&172.9&3.0&30.0&0.16&MORP\\
640315&0.417&0.900&12.3&175.0&105.4&355.0&15Mar&179.0&-10.6&31.0&0.18&PN\\
660313&0.299&0.770&5.2&172.0&118.3&352.0&12Mar&183.0&-6.6&26.6&0.18&PN\\
690318&0.39&0.840&2.8&177.0&109.4&357.0&17Mar&185.0&-5.0&28.7&0.10&PN\\
2007CA19&0.483&0.827&9.6&174.8&97.6&-&-&-&-&-&0.15&JPL\\
\hline
{\bf NDS}&{\bf 0.430}&{\bf 0.807}&{\bf 4.8}&{\bf 206.7}&{\bf 74.0}&{\bf 206.7}&{\bf 20Oct}&{\bf 199.3}&{\bf -3.1}&{\bf 27.1}&-&{\bf tw}\\
29 meteors&0.484&0.827&6.6&189.3&82.3&189.3&02Oct&187.5&5.2&26.9&0.15&MODC\\
\hline
{\bf SDS}&{\bf 0.459}&{\bf 0.794}&{\bf 3.3}&{\bf 23.8}&{\bf 257.0}&{\bf 203.8}&{\bf 17Oct}&{\bf 194.9}&{\bf -10.0}&{\bf 26.1}&-&{\bf tw}\\
240/DFV&0.525&0.650&2.6&22.3&258.1&202.3&15Oct&193.7&-9.6&21.1&0.16&S3\\
22 meteors&0.480&0.828&5.3&11.2&262.1&191.2&04Oct&183.9&-8.5&26.8&0.12&MODC\\
\hline
\end{tabular}
\end{flushleft}
\label{table5}}
\end{table*}

\begin{table*}
\caption{ Meteor showers, fireballs/meteors, and NEAs associated with NEA 2010 FJ81 (J2000.0).}
\scriptsize{
\begin{flushleft}
\begin{tabular}{lcccccccccccl}
\hline
Objects&$q$&$e$&$i^{\circ}$&$\Omega^{\circ}$&$\omega^{\circ}$&$L_{\odot}^{\circ}$&Date&$\alpha_{g}^{\circ}$&$\delta_{g}^{\circ}$&$V_{g}$&$D$&Ref.\\ 
&AU&&&&&&&&&km/s&$_{SH}$&\\
\hline
{\bf NNS}&{\bf 0.396}&{\bf 0.810}&{\bf 4.0}&{\bf 342.1}&{\bf 290.5}&{\bf 342.1}&{\bf 2Mar}&{\bf 173.9}&{\bf 6.6}&{\bf 27.6}&-&{\bf tw}\\
11/EVI&0.464&0.812&5.4&355.7&281.0&355.7&16Mar&184.8&3.9&26.6&0.09&J1\\
N.$\eta$-Virds&0.495&0.707&3.7&357.0&282.4&357.0&17Mar&185.1&2.6&23.0&0.17&S2\\
N.$\eta$-Virds&0.502&0.703&11.3&349.3&281.8&349.3&09Mar&182.3&13.6&23.5&0.19&S3\\
Virds&0.365&0.914&3.8&348.5&286.5&348.5&09Mar&178.1&4.4&31.0&0.11&G\\
N.$\eta$-Virds&0.377&0.831&3.9&352.6&291.5&352.6&13Mar&184.0&2.0&28.7&0.16&T1\\
As.68&0.346&0.820&5.5&331.6&296.8&331.6&20Feb&167.6&10.2&28.9&0.08&L\\
As.112&0.433&0.870&6.1&355.5&285.6&355.5&16Mar&185.3&2.6&29.6&0.14&L\\
680306&0.389&0.860&3.7&345.0&287.4&345.0&05Mar&176.0&5.3&29.3&0.05&PN\\
710319&0.537&0.71&5.6&358.0&276.9&358.0&18Mar&184.0&5.7&22.6&0.17&PN\\
140378&0.390&0.870&1.3&352.8&287.9&352.8&13Mar&182.0&0.53&29.7&0.14&EN\\
2003FB5&0.533&0.788&5.3&358.4&288.4&-&-&-&-&-&0.20&JPL\\
\hline
{\bf SNS}&{\bf 0.421}&{\bf 0.798}&{\bf 5.7}&{\bf 164.7}&{\bf 108.0}&{\bf 344.7}&{\bf 5Mar}&{\bf 171.6}&{\bf -2.4}&{\bf 26.9}&-&{\bf tw}\\
S.$\eta$-Virds&0.566&0.738&6.1&181.3&91.2&1.3&22Mar&179.1&-8.2&22.9&0.16&S2\\
S.$\eta$-Virds&0.499&0.706&2.6&170.4&101.8&350.4&11Mar&175.4&-1.4&22.9&0.13&S3\\
As.2&0.455&0.820&6.0&175.0&100.0&355.0&15Mar&177.0&5.0&26.7&0.05&K\\
S.$\eta$-Virds&0.397&0.904&12.3&175.2&105.4&355.2&15Mar&179.3&-10.8&31.2&0.19&T\\
790303&0.366&0.865&0.2&162.8&111.1&342.8&03Mar&172.9&3.0&30.0&0.13&MORP\\
640315&0.417&0.900&12.3&175.0&105.4&355.0&15Mar&179.0&-10.6&31.0&0.19&PN\\
680225&0.360&0.820&15.5&156.0&113.5&336.0&24Feb&165.0&-7.5&28.9&0.19&PN\\
690318&0.390&0.840&2.8&177.0&109.4&357.0&17Mar&185.0&-5.0&28.7&0.20&PN\\
750228&0.415&0.830&0.4&159.0&107.1&339.0&27Feb&168.5&4.6&28.0&0.13&PN\\
2007CA19&0.483&0.827&9.6&174.8&97.6&-&-&-&-&-&0.10&JPL\\
\hline
{\bf NDS}&{\bf 0.423}&{\bf 0.796}&{\bf 5.5}&{\bf 200.2}&{\bf 72.4}&{\bf 200.2}&{\bf 13Oct}&{\bf 192.9}&{\bf 0.3}&{\bf 26.9}&-&{\bf tw}\\
29 meteors&0.484&0.827&6.6&189.3&82.3&189.3&02Oct&187.5&5.2&26.9&0.07&MODC\\
\hline
{\bf SDS}&{\bf 0.397}&{\bf 0.809}&{\bf 4.0}&{\bf 23.0}&{\bf 249.6}&{\bf 203.0}&{\bf 16Oct}&{\bf 190.9}&{\bf -8.7}&{\bf 27.6}&-&{\bf tw}\\
240/DFV&0.525&0.650&2.6&22.3&258.1&202.4&15Oct&193.7&-9.6&21.1&0.20&S3\\
22 meteors&0.480&0.828&5.3&11.2&262.1&191.2&04Oct&183.9&-8.5&26.8&0.09&MODC\\
\hline
\end{tabular}
\end{flushleft}
\label{table6}}
\end{table*}

As seen from Tables 5 and 6, a mutual link of asteroids is also revealed. The present-day orbits of 2003 FB5 and 1995 EK1 correspond to the nighttime shower Northern $eta$-Virginids while the orbit of potentially hazardous asteroid 2007 CA19 is close to the nighttime shower Southern $eta$-Virginids. Moreover, their possible entering into the Earth's atmosphere will be described by the features of this showers. For instance, it is theoretically predicted that the period of presumed impact of 2007 CA19 is around of 5 March, the geocentric velocity is within 27-30 km/s, and the equatorial coordinates of the point from which it will be directed to the Earth are within  $\alpha_{g}$=170-180$^{\circ}$ and $\delta_{g}$=5- $-10^{\circ}$. Additionally, a lot of fireballs consistent with the nighttime Northern and Southern $eta$-Virginids showers confirms it activity.

It may be concluded that asteroids under study have a related association with the meteoroid stream producing four active meteor showers. Found relations and the comet-like orbits point to a cometary origin of these objects which might be considered as the parent bodies of the stream or large-sized extinct fragments of the comet-progenitor of the stream. 

\subsection{Asteroids associated with the $nu$-Virginid stream}

Association between the stream generating the Northern and Southern $nu$-Virginids and 5 asteroids listed in the third section of Table 2 is found. As a result of a search the theoretical nighttime showers of the asteroids were identified with these showers found by \citet{ter66,ter89}. The shower 124/SVI (EDM1, \citet{rud15}) very likely is consistent with the Southern $nu$-Virginids shower. The results for asteroids 2013 CU82 and 2013 TR13 are given in Tables~\ref{table7}  and ~\ref{table8} where all designations are the same as in Tables 3 and 4. For the rest asteroids of this set the outcomes are identical. Note, the connection of asteroid 2004 CK29 with this stream was previously revealed by \citet{bab12}.

\begin{table*}
\caption{ Meteor showers, fireballs, and NEAs associated with NEA 2013 CU82 (J2000.0).}
\scriptsize{
\begin{flushleft}
\begin{tabular}{lcccccccccccl}
\hline
Objects&$q$&$e$&$i^{\circ}$&$\Omega^{\circ}$&$\omega^{\circ}$&$L_{\odot}^{\circ}$&Date&$\alpha_{g}^{\circ}$&$\delta_{g}^{\circ}$&$V_{g}$&$D$&Ref.\\ 
&AU&&&&&&&&&km/s&$_{SH}$&\\
\hline
{\bf NNS}&{\bf 0.555}&{\bf 0.777}&{\bf 9.05}&{\bf 350.2}&{\bf 270.7}&{\bf 350.2}&{\bf 10Mar}&{\bf 176.8}&{\bf 16.5}&{\bf 24.3}&-&{\bf tw}\\ 
N.$\nu$-Virds&0.394&0.822&5.6&335.9&289.6&335.9&24Feb&167.4&11.2&27.6&0.17&T\\
N.$\nu$-Virds&0.521&0.790&2.3&348.8&274.4&348.8&9Mar&173.0&6.0&24.8&0.13&T1\\ 
710319&0.537&0.710&5.6&358.0&276.9&358.0&18Mar&184.0&5.7&22.6&0.20&PN\\
\hline
{\bf SNS}&{\bf 0.582}&{\bf 0.766}&{\bf 9.59}&{\bf 173.0}&{\bf 88.0}&{\bf 353.0}&{\bf 13Mar}&{\bf 168.7}&{\bf -8.6}&{\bf 23.5}&-&{\bf tw}\\
124/SVI&0.576&0.753&2.8&170.6&88.5&350.6&11Mar&169.1&-1.0&22.5&0.12&EDM1\\ 
680330&0.684&0.870&3.9&189.0&71.3&9.0&29Mar&177.0&-5.4&22.6&0.18&PN\\ 
750321&0.710&0.680&6.7&180.0&73.9&0.0&20Mar&168.5&-7.7&18.8&0.19&PN\\ 
\hline
{\bf NDS}&{\bf 0.582}&{\bf 0.766}&{\bf 9.59}&{\bf 173.0}&{\bf 88.0}&{\bf 173.0}&{\bf 15Sep}&{\bf 174.8}&{\bf 19.7}&{\bf 23.5}&-&{\bf tw}\\
2002BK25&0.576&0.749&11.9&156.5&103.6&156.5&29Aug&163.8&23.5&23.1&0.06&JPL,tw\\
\hline
{\bf SDS}&{\bf 0.555}&{\bf 0.777}&{\bf 9.0}&{\bf 351.8}&{\bf 269.1}&{\bf 171.8}&{\bf 14Sep}&{\bf 166.9}&{\bf -6.30}&{\bf 24.2}&-&{\bf tw}\\
 2002TZ59&0.410&0.842&16.5&7.4&249.7&187.4&1Oct&172.0&-12.4&29.3&0.22&JPL,tw\\
2004CK39&0.393&0.829&12.3&359.5&265.1&-&-&-&-&-&0.19&JPL\\
\hline
\end{tabular}
\end{flushleft}
\label{table7}}
\end{table*}

\begin{table*}
\caption{ Meteor showers, fireballs/meteors, and NEAs associated with NEA 2013 TR13 (J2000.0).}
\scriptsize{
\begin{flushleft}
\begin{tabular}{lcccccccccccl}
\hline
Objects&$q$&$e$&$i^{\circ}$&$\Omega^{\circ}$&$\omega^{\circ}$&$L_{\odot}^{\circ}$&Date&$\alpha_{g}^{\circ}$&$\delta_{g}^{\circ}$&$V_{g}$&$D$&Ref.\\ 
&AU&&&&&&&&&km/s&$_{SH}$&\\
\hline
{\bf NNS}&{\bf 0.439}&{\bf 0.803}&{\bf 0.3}&{\bf 336.4}&{\bf 285.0}&{\bf 336.4}&{\bf 25Feb}&{\bf 164.9}&{\bf 6.8}&{\bf 26.6}&-&{\bf tw}\\
N.$\nu$-Virds&0.394&0.822&5.6&335.9&289.6&335.9&24Feb&167.4&11.2&27.6&0.11&T\\
N.$\nu$-Virds&0.521&0.79&2.3&348.8&274.4&348.8&09Mar&173.0&6.0&24.8&0.09&T1\\
As.68&0.346&0.820&5.5&331.6&296.8&331.6&20Feb&167.6&10.2&28.9&0.16&L\\
800222&0.498&0.754&2.8&332.5&278.9&332.5&21Feb&160.2&11.9&24.2&0.16&MORP\\
810210&0.355&0.822&4.4&320.9&294.8&320.9&09Feb&156.3&13.9&28.6&0.14&MORP\\
680306&0.389&0.86&3.7&345.0&287.4&345.0&05Mar&176.0&5.3&29.3&0.19&PN\\
2013CU82&0.563&0.744&10.8&317.7&303.2&-&-&-&-&-&0.20&JPL\\
\hline
{\bf SNS}&{\bf 0.474}&{\bf 0.787}&{\bf 2.1}&{\bf 159.7}&{\bf 101.7}&{\bf 339.7}&{\bf 28Feb}&{\bf 165.3}&{\bf 3.9}&{\bf 25.6}&-&{\bf tw}\\
S.$\nu$-Virds&0.387&0.85&0.3&160.4&109.1&340.4&01Mar&170.0&4.0&28.5&0.16&T\\
As.65&0.427&0.85&1.6&150.6&104.3&330.6&19Feb&158.8&7.3&28.4&0.12&L\\
840223&0.515&0.733&8.4&153.2&97.6&333.2&21Feb&155.9&-7.0&23.7&0.19&MORP\\
750228&0.415&0.83&0.4&159.0&107.1&339.0&27Feb&168.5&4.6&28.0&0.10&PN\\
\hline
{\bf NDS}&{\bf 0.476}&{\bf 0.786}&{\bf 1.9}&{\bf 182.4}&{\bf 79.0}&{\bf 182.4}&{\bf 25Sep}&{\bf 178.2}&{\bf 3.0}&{\bf 25.5}&-&{\bf tw}\\
2002BK25&0.576&0.749&11.9&156.5&103.6&156.5&29Aug&163.8&23.5&23.1&0.21&JPL,tw\\
\hline
{\bf SDS}&{\bf 0.439}&{\bf 0.803}&{\bf 0.3}&{\bf 5.8}&{\bf 255.6}&{\bf 185.8}&{\bf 29Sep}&{\bf 178.6}&{\bf 0.3}&{\bf 26.7}&-&{\bf tw}\\
24 meteors&0.426&0.798&8.2&12.6&243.9&192.6&05Oct&179.6&-8.0&28.4&0.15&MODC\\
2002TZ59&0.41&0.842&16.5&7.4&249.7&187.4&1Oct&172.0&-12.4&29.3&0.20&JPL,tw\\
2004CK39&0.393&0.829&12.3&359.5&265.1&-&-&-&-&-&0.20&JPL\\
\hline
\end{tabular}
\end{flushleft}
\label{table8}}
\end{table*}

The theoretically predicted northern branch of the daytime shower doesn't identify with any known shower. The southern branch of the daytime shower was identified with 24 meteors from the MODC database, their mean radiants positions and the activity dates allow classifying them as the Daytime October $beta$-Virginids (Table 8).  As the theoretical daytime shower of the asteroids belonging to the $eta$ -Virginids family were also identified with the Daytime October $beta$-Virginids it may be concluded that the Virginid complex in fact includes a lot of close meteor showers and subshowers. 

As seen, a mutual association of asteroids is also established. The present-day orbit of 2013 CU82 corresponds to the northern branch of nighttime shower, the potentially hazardous asteroid 2002 BK25 relates to the northern branch of daytime shower, and asteroids 2002 TZ59 and 2004 CK39 match to the southern branch of daytime shower. Presumed impacts of these asteroids will have the features similar to who's of the relevant showers, namely, the identical dates, velocities and the coordinates of positions from which the entering will be realized. Particularly, for the potentially hazardous asteroid 2002BK25 the coordinates of the position from which it will be directed to the Earth are $\alpha_{g}$=163.8$^{\circ}$ and $\delta_{g}$=23.5$^{\circ}$, the geocentric velocity is 23.1 km/s, and the date is 30 August.

Found association with the stream and active showers confirms that asteroids under consideration having the comet-like orbits have produced meteoroids, from which the stream was formed, during very long period, hence they might be at present extinct comets.

\subsection{Asteroids associated with the $mu$-Virginid stream}

It was defined that four asteroids given in the fourth section of Table 2 relate to the meteoroid stream producing the nighttime Northern and Southern $mu$-Virginids \citep{coo73,sek73,sek76,ter66,ter89}. The results of identification of the theoretical and observable showers, fireballs, and NEAs for asteroids 2010 XD11 and 2010 GO33 are given in Tables~\ref{table9} and ~\ref{table10} where all designations are the same as in Tables 3 and 4. For the rest asteroids of this set the outcomes are identical.

\begin{table*}
\caption {Meteor showers, fireballs, and NEAs associated with NEA 2010 XD11 (J2000.0).}
\scriptsize{
\begin{flushleft}
\begin{tabular}{lcccccccccccl}
\hline
Objects&$q$&$e$&$i^{\circ}$&$\Omega^{\circ}$&$\omega^{\circ}$&$L_{\odot}^{\circ}$&Date&$\alpha_{g}^{\circ}$&$\delta_{g}^{\circ}$&$V_{g}$&$D$&Ref.\\ 
&AU&&&&&&&&&km/s&$_{SH}$&\\
\hline
{\bf NNS}&{\bf 0.356}&{\bf 0.839}&{\bf 5.3}&{\bf 19.8}&{\bf 294.4}&{\bf 19.8}&{\bf 09Apr}&{\bf 210.6}&{\bf -7.8}&{\bf 29.3}&-&{\bf tw}\\
N.$\mu$-Virds&0.530&0.830&10.0&35.0&280.0&35.0&25Apr&221.0&-5.0&29.0&0.20&C\\
N.Virds&0.278&0.785&4.8&11.9&310.3&11.9&01Apr&210.4&-8.5&27.4&0.15&S3\\
N.$\mu$-Virds&0.357&0.885&2.2&20.5&291.7&20.5&10Apr&209.0&-10.0&30.8&0.08&T1\\
220495&0.499&0.790&4.1&31.7&277.5&31.7&21Apr&215.0&-8.9&25.1&0.17&EN\\
680330&0.270&0.900&9.5&9.0&302.4&9.0&29Mar&206.0&-3.9&33.4&0.14&PN\\
\hline
{\bf SNS}&{\bf 0.325}&{\bf 0.853}&{\bf 5.5}&{\bf 196.4}&{\bf 117.8}&{\bf 16.4}&{\bf 06Apr}&{\bf 205.8}&{\bf -15.1}&{\bf 30.3}&-&{\bf tw}\\
651/OAV&0.373&0.842&4.7&197.4&109.8&17.4&07Apr&202.0&-13.5&28.9&0.11&J1\\
(S.$\mu$-Virds?)&&&&&&&&&&&&\\
S.Virds&0.312&0.828&0.0&183.7&121.8&3.7&24Mar&197.4&-7.4&29.2&0.16&S2\\
S.$\mu$-Virds&0.264&0.864&3.5&194.5&126.0&14.5&04Apr&208.0&-14.0&31.5&0.12&T1\\
865&0.373&0.715&1.2&193.5&120.3&13.5&03Apr&205.4&-11.8&24.2&0.16&MORP\\
1011&0.256&0.949&5.8&188.0&122.1&8.0&28Mar&299.5&-11.9&35.8&0.13&MORP\\
680411&0.375&0.850&4.9&201.0&109.4&21.0&11Apr&205.0&-15.7&28.9&0.08&PN\\
\hline
{\bf NDS}&{\bf 0.329}&{\bf 0.851}&{\bf 5.2}&{\bf 251.5}&{\bf 62.7}&{\bf 251.5}&{\bf 03Dec}&{\bf 237.4}&{\bf -15.9}&{\bf 30.1}&-&{\bf tw}\\
-&\multicolumn{12}{c}{Observable showers/objects not found}\\
\hline
{\bf SDS}&{\bf 0.356}&{\bf 0.839}&{\bf 5.3}&{\bf 68.7}&{\bf 245.5}&{\bf 248.7}&{\bf 01Dec}&{\bf 233.8}&{\bf -23.9}&{\bf 29.2}&-&{\bf tw}\\
2003WW26&0.482&0.798&6.3&57.7&254.6&237.7&20Nov&228.8&-25.2&25.7&0.14&JPL,tw\\
 \hline
\end{tabular}
\end{flushleft}
\label{table9}}
\end{table*}

\begin{table*}
\caption { Meteor showers, fireballs, and NEAs associated with NEA 2010 GO33 (J2000.0).}
\scriptsize{
\begin{flushleft}
\begin{tabular}{lcccccccccccl}
\hline
Objects&$q$&$e$&$i^{\circ}$&$\Omega^{\circ}$&$\omega^{\circ}$&$L_{\odot}^{\circ}$&Date&$\alpha_{g}^{\circ}$&$\delta_{g}^{\circ}$&$V_{g}$&$D$&Ref.\\ 
&AU&&&&&&&&&km/s&$_{SH}$&\\
\hline
{\bf NNS}&{\bf 0.713}&{\bf 0.703}&{\bf 10.5}&{\bf 45.4}&{\bf 251.8}&{\bf 45.4}&{\bf 6May}&{\bf 220.8}&{\bf 2.6}&{\bf 19.6}&-&{\bf tw}\\
N.$\mu$-Virds&0.696&0.748&3.3&43.2&253.7&43.2&3 May&214.0&-8.0&20.1&0.13&T1\\
454/MPV&0.650&0.744&10.4&41.6&259.7&41.6&2May&220.2&0.3&21.7&0.09&EDM2\\
(N.$\mu$-Virds?)&&&&&&&&&&&&\\
307&0.660&0.563&11.9&40.5&267.4&40.5&30Apr&224.7&4.0&18.0&0.19&MORP\\
730503&0.636&0.630&9.6&42.0&267.5&42.0&2May&224.0&-1.4&19.4&0.17&PN\\
\hline
{\bf SNS}&{\bf 0.680}&{\bf 0.717}&{\bf 10.6}&{\bf 221.0}&{\bf 76.2}&{\bf 41.0}&{\bf 1May}&{\bf 207.1}&{\bf -28.8}&{\bf 20.6}&-&{\bf tw}\\
S.$\mu$-Virds&0.720&0.696&2.0&223.1&72.3&43.1&3May&210.0&-16.0&18.6&0.16&T1\\
690512&0.800&0.680&6.0&231.0&63.5&51.0&11May&210.0&-25.0&17.1&0.15&PN\\
2011VG9&0.508&0.776&1.3&236.1&65.8&-&-&-&-&-&0.25&JPL\\
\hline
{\bf NDS}&{\bf 0.675}&{\bf 0.719}&{\bf 10.6}&{\bf 194.2}&{\bf 103.0}&{\bf 194.2}&{\bf 7Oct}&{\bf 204.0}&{\bf 7.8}&{\bf 20.8}&-&{\bf tw}\\
  -&\multicolumn{12}{c}{Observable showers/objects not found}\\
\hline
{\bf SDS}&{\bf 0.716}&{\bf 0.703}&{\bf 10.5}&{\bf 9.3}&{\bf 288.0}&{\bf 189.3}&{\bf 3Oct}&{\bf 188.5}&{\bf -22.9}&{\bf 19.6}&-&{\bf tw}\\
-&\multicolumn{12}{c}{Observable showers/objects not found}\\
\hline
\end{tabular}
\end{flushleft}
\label{table10}}
\end{table*}

As was mentioned, $mu$-Virginids is denoted in the IAU MDC as 47/DLI (\citet{por94} (P)). However, according to the radiant position the shower should be named as $delta$-Librids, moreover the abbreviation DLI also points to this definition. While the $mu$-Virginids should be named as MVI. The radiants and orbits of the 47/DLI given by \citet{rud15} (EDM1)  and \citet{jen16} (J1)  are more consistent with the $xi$1-Librid shower. By this reason, although the 47/DLI was identified with the theoretical shower, three references P, EDM1 and J1 are excluded from the Tables 9 and 10. The observable October $alpha$-Virginid shower (651/OAV, \citet{jen16} (J1)) was identified with the southern branch of theoretical night-time shower of asteroid 2010 XD11 (Table 9), however the shower should not be classified as October since it value of the Solar longitude corresponds to 7 April. So that we refer the shower 651/OAV to the Southern $mu$-Virginid shower.
   
A mutual relationship of the asteroids also is found. The present-day orbit of 2011 VG9 corresponds to the southern branch of nighttime shower, and the orbit of 2003 WW26 - to the southern branch of daytime shower. At the case of possible impacts with the Earth, for example for 2003 WW26, the coordinates of the point of entering into the Earth’s atmosphere will be $\alpha_{g}$ =228.8$^{\circ}$ and $\delta_{g}$=-25.2$^{\circ}$,  the geocentric velocity $V_{g}$=25.7 km/s and the date - 20 November.

Theoretical daytime showers were not identified with observables, because showers with such features are yet missed in the used databases of observations. However, identification of the theoretical nighttime showers allows suggesting the related association of the asteroids with the stream and these showers. The asteroids very likely have a cometary nature and might be extinct fragments of the parent comet of the $mu$-Virginids meteoroid stream.

\section{Discussion}

Thus, the existence of \, \, active \, \, meteor showers \, \, producing by \, \, the \, \, $alpha-, eta-, nu-$ and $mu$-Virginid meteoroid streams, and their relationship with four groups of asteroids are found. These streams and substreams, and generating by them meteor showers relate to the Virginid complex. Revealed association is very strong indicator of a cometary origin of 29 NEAs moving on the comet-like orbits. In this case and taking into account that each group contains a few of asteroids several scenario of their formation might be suggested. 1. The largest object is the extinct parent Jupiter family comet of the stream, and the rest objects are its fragments also extinct at present. 2. All objects of each group were formed as a result of a decay of the larger comet-progenitor of the relevant stream. 3. As a result of a break-up of the one giant parent comet of the Virginid complex smaller comets were produced which during evolution under gravitational and non-gravitational perturbations generated the substreams of the Virginid complex. These inactive comets at present compose four groups under consideration.

However, whatever the mechanism of formation, this leads to an unambiguous conclusion, namely, the Virginid complex includes four families that we identified. Each of them consists of the meteoroid stream generating active showers and contains from 14 ($alpha$-Virginids stream) to 4 ($mu$-Virginid substream) the large-sized extinct remnants of parent comets (or comet). Besides, found association with the known meteorites points to that the large-sized objects having more early asteroidal origin should be in the stream which are capable to produce meteorites in future. The asteroids of a cometary nature prove that the large bodies with the size from 110 m to 1.2 km are the constituent part of the stream. Asteroid 2002 BK25 with the diameter of 1.22 km is already classified as the potentially hazardous object in this complex.

\section{Conclusions}
A new Virginids asteroid-meteoroid complex is established as a result of performed investigation. The complex includes several meteoroid streams and substreams producing the active meteor showers confirmed by the observations. Besides, the related association of 29 near-Earth's asteroids with the complex is found. This and the comet-like orbits allow suggesting that asteroids with high probability have a cometary origin. In this case they might be considered as the fragments of a parent comet of the Virginid complex which are presently in extinct or dormant stage. 

 The presented study again confirm that the meteoroid streams consist both of small mm sized particles and large bodies of decameter size and more  which represent potentially hazard for the Earth and the space missile technology. Our results allow forecasting the parameters of an incoming of such bodies into the Earth's atmosphere that is need for a development of the mitigation strategy for preventing the consequences.  

The results show that the method of identification of extinct comets among the near-Earth's asteroids is effective and yields the reliable outcomes. For more sure additional observational data for the Virginid complex showers need in order to obtain the dynamical and physical properties of meteoroids of the streams and substreams of this complex which is associated with 29 objects of a cometary nature. 

A search for other presumed existent large fragments of parent comets should be continued in future. They also presently are extinct and might be found among both known NEAs and new discovered ones.

\end{document}